# A Relativistic DFT Study of Water Adsorption on δ-Plutonium (111) Surface

Raymond Atta-Fynn and Asok K. Ray*

*Department of Physics, University of Texas at Arlington, Arlington, Texas 76019*

*akr@uta.edu

## Abstract

Scalar-relativistic DFT-GGA has been used to study adsorption of water in molecular and dissociative configurations on δ-Pu (111) surface. In molecular state, water is physisorbed in an almost flat-lying orientation at a one-fold coordinated on-top site. The interaction of the water $1b_1$ orbital and the Pu-$6d$ orbital provides the stability of water on the surface, implying that the Pu-$5f$ electrons remain chemically inert. The co-adsorption cases of partially dissociated and fully dissociated products at the three-fold hollow sites yield chemisorption, coupled with rumpling of the surface layer and delocalization of the Pu-$5f$ electrons and formation of strong ionic bonds.

1. Introduction

The actinides, a group of fourteen radioactive and electropositive elements with atomic numbers 90 to 103 that follow Actinium in the periodic table, are characterized by the gradual filling of the valence 5$f$ shells [1-6]. Plutonium (Pu) occupies a crucial position in the actinide series as elements to the left (thorium (Th) to neptunium (Np)) have itinerant 5$f$ electrons and the elements to its right (americium (Am) and beyond) have localized 5$f$ electrons. Thus Pu is the point where 5$f$ electrons delocalization to localization transition in the actinides occurs. As a result, Pu is a chemically complex metal with six different crystal structures between room temperature and an anomalously low melting point of 913°K under atmospheric conditions. Among all the different Pu crystal structures, δ-Pu is most technologically important because it is ductile and malleable (the ground state α-Pu is hard and brittle) and can be stabilized at room temperature with a fcc structure when alloyed with a few atomic percent of impurities such as Ga and Al. Controversies abound in the literature concerning the electronic structure of δ-Pu, specifically regarding the magnetic nature and the localization/delocalization behavior of its 5$f$ electrons. One way to probe this problem is a study of the δ-Pu surface and its reactivity with atomic and molecular complexes, a rather important example being the $H_2O$ molecule. The water molecule is an excellent probe for studying substrate properties such as catalytic reactivity, redox processes, site distributions, defect influence, adsorption capacity and electronic corrugation among others [7].

Despite the obvious importance, there are very few studies in the literature concerning the surface reactivity of δ-Pu. Haschke *et al.* [8] studied the reaction kinetics of unalloyed Pu with oxygen, water, and observed that the corrosion rate of water increases steadily from 35-110 °C with an activation energy of 1.214± 0.03 eV and in the range of 110-200 °C, the rate decreases with an activation energy of 0.65 to -0.867 eV. Boettger and Ray [9], Wu and Ray [10] and Hay [11] studied the interactions of water molecule with $PuO_2$ surfaces using DFT-GGA model. Dissociative adsorption was found to be preferred over molecular adsorption. In this Letter, we report DFT-based calculations on the adsorption and dissociation of water on the (111) surface of δ-Pu. Our goal is to investigate the preferred adsorbate configuration, the nature of the binding of the adsorbate to the surface and the adsorbate-induced changes in the surface structural and electronic properties.

## 2. Computational details

DFT-GGA calculations using the Perdew-Burke-Enzerhof (PBE) [12] formulation have been carried using the codes $DMol^3$ [13, 14] and WIEN2k [15]. The DMol3 code uses numerical basis sets, and both pseudo-potential and all-electron scalar relativistic calculations can be performed. WIEN2k is an all-electron code which employs a full-potential linearized augmented plane wave plus local basis (FP-L/APW+lo) method, and both scalar-relativistic and spin-orbit-coupling inclusive calculations can be performed. First, full structural optimizations were performed with $DMol^3$ because of computational efficiency. The resulting optimized structures were then used as inputs in WIEN2k (single-point scalar relativistic energy runs) because of its accuracy in predicting local electronic and magnetic properties.

For the $DMol^3$ calculations, the expansion of the single particle Kohn-Sham orbital was carried out using a double numerical basis set with polarization functions (DNP) and a real space cut-off of 5.5 Å. The hardness conserving density functional semicore pseudopotential (DSPP), with the semicore states being treated as valence states, provided with the code was used. The bulk (f.c.c) unit cell was first optimized, resulting in an anti-ferromagnetic (AFM) ground state with a lattice constant of 8.70 a.u., a 0.7 % contraction of the experimental lattice constant of 8.76 a.u [1]. The surface, with a *p* (2x2) surface unit cell, was then modeled by repeated slabs with five atomic layers separated by a vacuum of 30 a.u. thickness. A full relaxation of the δ-Pu (111) slab yielded interlayer relaxations of $\Delta_{12}$=3.0% between the surface and subsurface layers and $\Delta_{23}$=0.0 % between the subsurface and bulk layers. The DMol$^3$ optimizations were terminated when the self-consistent density, total energy, atomic forces, and atomic displacements converged to within $5\times10^{-4}$, $10^{-5}$ Ha, $4\times10^{-3}$ Ha/Å, and $3\times10^{-3}$ Å, respectively.

For the WIEN2k FP-LAPW+lo calculations, the unit cell is divided into non-overlapping muffin tins and an interstitial region. Inside the muffin tins, the Kohn-Sham orbital is expanded using radial functions multiplied by spherical harmonics up to $l_{max}$ =10, while plane waves with kinetic energies up to 16 Ry were used for the expansion in the interstitial region. The muffin tin radii used were $R_{MT}$ (Pu) = 2.13 a.u., $R_{MT}$ (O) = 1.2 a.u., and $R_{MT}$ (H) = 0.6 a.u., keeping in mind that the choice of the radii values depends on the specific problem being studied under the condition that the muffin-tin spheres do not overlap. Additional local orbitals were added to the Pu 6s, 6p and O 2p semi-core states to improve their description. For the WIEN2k calculations, the self-consistent charge density and total energy convergence of $10^{-3}$ and $10^{-5}$ Ry respectively were employed. The Brillouin zone (BZ) was sampled by a 5 × 5 × 1 mesh of *k* points. A thermal broadening of the Fermi surface corresponding to $K_B T=0.01$ Ry was employed to accelerate the convergence of the electronic density. The standard procedure for including scalar relativistic corrections in both WIEN2k and DMol$^3$ is due to the approximation developed by Koelling and Harmon [16] to include the mass-velocity and Darwin s-shift but omits the spin-orbit terms by performing a *j*-weighted average of the Dirac equations for a spherically symmetric potential field $V(r)$.

To study adsorption on the surface, we considered four high symmetry adsorption sites on the f.c.c. (111) surface: i) one-fold on-top site $T_1$; ii) two-fold bridge site $B_2$; iii) three-fold hollow hcp site $H_3$; and iv) three-fold hollow fcc site $F_3$ (Fig.1). The adsorbates were initially placed on both faces of the fully relaxed slab to preserve inversion symmetry and the entire system was again relaxed. The computations of the free energies of the adsorbates and the structural optimization of the molecules were carried out in the same supercell used for the slab calculation with the same computational parameters. For the $H_2O$ molecule, the equilibrium OH bond length and HOH bond angle were found to be 0.97 Å and 103.6°, respectively, the experimental values being 0.96 Å and 104.5° respectively [17]. The adsorption energy (per adsorbate) is given by

$$E_{ads} = \frac{1}{2N}\left[E(\text{slab}) + 2E(\text{adsorbates}) - E(\text{slab}+\text{adsorbates})\right],$$

where E is the total energy of the species in parentheses and N is the number of adsorbates. As mentioned before, all computations were carried out at the scalar relativistic level. SOC-inclusive adsorption energies were not considered because our previous works indicate that on the average, SO changes the adsorption energy by about a tenth of an eV, partly due to cancellation in the formula above [18].

## 3. Results and discussions

We have listed in Table I the adsorption energies and the equilibrium geometric parameters of the adsorbates on the δ-Pu (111) surface. For each molecular adsorption site, the molecule was initially placed in three orientations relative to the surface: α=0° (lying flat and parallel to the surface), α=90° (upright and perpendicular to the surface with H pointing up), and α= -90° (upright and perpendicular to the surface with H pointing down). Placing the molecule on the surface for different initial values of α is necessary to avoid being trapped by possibly small activation barriers as much as possible. The adsorbates were allowed to relax from initially symmetric sites to other sites on the surface to search for a global minimum [19]. We observe that the adsorption process is possibly physisorption with the most stable configuration corresponding to an adsorption energy of E=0.58 to 0.59 eV, where all the molecules relax from their initial non-$T_1$ site to the neighboring $T_1$ site or stay at their initial $T_1$ site. Also, the molecule relaxes from their initial state (orientation) $α_i$ = -90°, 0°, 90° to a final state $α_f$ =9° to12°, which implies that $H_2O$ molecule prefers to adsorb on-top with nearly flat lying dipole moment relative to the surface. The next set of adsorption energies for configurations with $T_1$ as the final site, all of which lie in the range 0.53 eV to 0.57 eV and are thus within a few hundredths of eV from the most stable energies, also posses $α_f$ = -2° to 6°. This clearly points to the fact that the on-top $T_1$ site is the global minimum and that the three atoms of water together with the Pu atom at the $T_1$ site form a tetrahedron. However, in all cases, there is a non-zero lateral shift $\Delta O_{xy}$ of the oxygen atom from the precise symmetric $T_1$ site ranging from 0.09 Å to 0.79 Å. We wish to point out that the $H_2O$ adsorption geometry at the $T_1$ site has been observed for other close-packed transition and noble metal surfaces [19]. From Table I, we note that the next and only preferred final site after $T_1$ for molecular adsorption is $B_2$. The most stable $B_2$ configurations are tilted to angles of 52° and 61° with respect to the surface and therefore do not necessarily prefer a nearly-flat lying molecular orientation. Two least energetically favorable configurations correspond to the $B_2$ site with the molecule nearly upright and the H atoms pointing towards the surface. Overall the bond length $d_{OH}$ and bond angle θ of the molecule after adsorption show little or no departure from that of the free molecule. In all cases, the top layer expands by $\Delta_{12}$=1.4 % to 2.5 %, the magnitude of which is smaller than the bare layer relaxation of $\Delta_{12}$=3.0 %. Adsorption at the $T_1$ and $B_2$ sites results in one or more Pu atoms leaving the top metal layer and moving inwards or outwards. This causes the surface to rumple after adsorption. The amount of rumpling Δr is the maximum displacement between two atoms originally on the same atomic layer. The computed surface rumpling was Δr = 0.05 Å to 0.65 Å, for the most stable $T_1$. Also, the rather large nature of the interatomic distances $d_{OPu}$ and $d_{HPu}$ in Table 1 provides further basis that the $H_2O$ adsorption process is possibly physisorption. Furthermore, we used the Mulliken charges from the DMol calculations to gauge the amount of adsorbate-substrate transfer. We observed a maximum charge transfer of 0.04 $e$ from O to the Pu surface and 0.01 $e$ from the H atoms, which is also indicative of the possible physisorptive nature of the adsorption process. To test the validity of our calculations in regards to interactions between an adsorbate and its periodic counterpart in the next unit cell, we carried out calculations for water adsorbed at a coverage of Θ=1/9 using a $p$ (3×3) surface unit cell and compared with our current results for Θ=1/4 using a $p$ (2x2) cell. We considered two initial configurations at the $T_1$ site: $α_i$ = -90° and $α_i$ = 0° orientations. For the initial $α_i$ = 0° state, $ΔE_{ads}$ = 0.08 eV, $Δα_f$ = 4°, and $d_{OPu}$ = 0.04 Å. For the initial $α_i$ = -90° state, $ΔE_{ads}$ = 0.08 eV, $Δα_f$ = 3°, and $d_{OPu}$ = 0.03 Å. Hence the nearly flat geometry of the adsorbed water molecule is preserved at the top site at a lower coverage. Thus, it is reasonable to conclude that the interaction between an admolecule and its periodic image in the neighboring cell for $p(2×2)$ surface is sufficiently weak and does not affect the conclusions of our work.

Having established that the $T_1$ site is the most favorable site for $H_2O$ adsorption, we investigated the energy barrier that the molecule will encounter as it diffuses across the surface. Diffusion from a $T_1$ site to a neighboring $T_1$ site occurs via a $B_2$, $H_3$ or $F_3$ site. However from Table 1, it is obvious that among the three sites $B_2$, $H_3$ and $F_3$, the $B_2$ site is the most stable. Thus the lowest energy pathway for diffusion across the surface must be the $T_1$-$B_2$-$T_1$ route. In Figure 2, we depict the energy profile for the diffusion of the molecule along the $T_1$-$B_2$-$T_1$ pathway. Using seven displacements between the initial state (IS) and the final state (FS), and constraining the molecule to stay at each point on the path during relaxation, we estimate the energy barrier to diffusion $E_b$ to be 0.18 eV. The transition state (TS) is located at a point with a lateral separation of 0.23 Å from the $B_2$ site along the path. In fact the $B_2$ site corresponds to the point to the immediate left of TS in the figure. Usually on the fcc (111) surface, the energy barrier to diffusion is determined by simply taking the difference between the energies at IS and the bridge site since TS is located at the bridge site. However, our calculations indicate that the transition point is not precisely located at the $B_2$ site. A similar observation was made by Michaelides *et al*. [19], who studied the $T_1$-to-$T_1$ diffusion of water on the Al(100) surface using four different routes including the route used here.

Several initial configurations can be used to simulate the co-adsorption of the partially dissociated H and OH products, since each adsorbate can occupy four sites and the OH molecule can have several orientations. However, we can use the nature of the adsorption of the individual products on the surface as an *ad hoc* scheme to eliminate energetically unfavorable sites. Previous work has shown that H adsorbs on δ-Pu (111) with a preference for the $F_3$ and $H_3$ sites [20]. To determine the likely energetically favorable configurations, we adsorbed only OH on the surface at each of the four adsorption sites and in three initial orientations: (i) vertically upright with H up and O down ($α_i$ =90°) (ii) vertically upright with H down and O up ($α_i$ = -90°) (iii) flat-lying ($α_i$ = 0°). In all cases, the molecule relaxed to an $F_3$ or $H_3$ site in a $α_f$ =90° orientation or dissociates into an $α_f$ = -90° orientation at an $F_3$ or $H_3$ site with H moving to occupy the hollow adsorption site below the first layer. We thus inferred that the most energetically plausible configuration will most likely correspond to both H and OH at neighboring $F_3$ sites or neighboring $H_3$ sites. The results for the (OH+H) co-adsorption are presented in the second panel in Table 1. Clearly the binding energies are nearly degenerate with the molecule preferring a $α_f$ =90° orientation. The chemisorption energies of 5.16 eV to 5.19 eV are about an order of magnitude larger than the molecular physisorption energies. As can clearly be seen, the final orientation of hydroxyl was $α_f$ =90° in all four cases. Furthermore, each adsorbate stayed at the initial at or relax to the symmetric $F_3$ or $H_3$ adsorption site unlike $H_2O$ adsorption. The OH bond lengths contract by 0.02 Å- 0.03 Å and the top layer expands significantly compared to the bare surface except for one case where the expansion is less than that of the clean surface. The rearrangement of the surface Pu atoms to accommodate the adsorbates leads to surface rumplings of Δr = 0.20 Å to 0.93 Å. Coordination number plays an important role in the stabilization of chemisorbed system. Because the O-Pu and H-Pu bond lengths are within right bonding range, the more the adsorbate overlaps with neighboring Pu atoms the stronger the interaction- an observation which is certainly not true for the molecular case. Thus higher coordination number in this case implies stronger binding.

Following the discussion above, we expect the energetically favorable configuration for the co-adsorption of O and two H atoms to occur at neighboring $F_3$ or $H_3$ sites as previous studies have shown that both adsorbates adsorb favorably at $F_3$ or $H_3$, *albeit* with nearly degenerate chemisorptions energies [18, 20]. Thus the simulation of the fully dissociated products was carried out by placing each product at neighboring $F_3$ sites or at neighboring $H_3$ sites. In the third panel in Table 1, we report the chemisorptions energies and geometries of the adsorption systems. It is immediately evident that the adsorption energies for the fully dissociated product is more stable than that of the partially dissociated products, which in turn is more stable than that of the molecule. The average relaxation of the top layer is surprisingly small. However, the 1.0 Å and 1.5 Å. rumpling of the surface layer are quite significant. This is because at such a high coverage of 75%, the substrate must undergo significant structural rearrangements to accommodate the three adsorbates. The range of the bond distances and the three-fold coordination of each atom guarantees maximum adsorbate-substrate overlap and hence the binding is expected to be more stable. .

We also computed the electron difference charge density distribution Δ$n$(r) from Δ$n$(r) = $n$(adsorbate+slab) − $n$(slab) − $n$(adsorbate), where $n$(adsorbate+slab) is the total electron charge density of the δ-Pu(111) slab-with-adsorbate, $n$(slab) is the total charge density of the clean δ-Pu(111) slab, and $n$(adsorbate) is the total charge density of the adsorbate. In computing $n$(adsorbate) and $n$(slab), the atomic positions of the adsorbate and the clean slab corresponded to their respective positions in slab-with-adsorbate system. In Figure 3, we show the planar plots of Δ$n$(r) for the most stable configuration corresponding to the most stable adsorption of the molecular, partially dissociated, and fully dissociated products. In Figure 3(a), charge build-up in the region between the polarized O atom and the Pu atom below it can clearly be seen, with charge depletion around Pu and an oxygen π-like orbital oriented along the z-axis. The charge build-up between Pu and O implies that the O-Pu has a covalent character. On the other hand, the build-up is weighted more towards O, implying that there is a net charge transfer from the O π-like orbital towards the surface, and thus O-Pu bond has a small ionic contribution. From the difference charge density plots for the $H_2O$/δ-Pu(111) system, we note that the O π-orbital, a 1$b_1$ orbital, is responsible for the stabilization of $H_2O$ on δ-Pu(111) and that the orientation of the induced dipole moment due to charge re-arrangements points out of the surface.

In Figures 3(b) and 3(c), the difference charge density plots for (OH+H)/δ-Pu(111) system and (O+2H)/δ-Pu(111) systems are shown. Clearly the O-Pu and H-Pu bonds in both figures are mainly ionic in character since there is a significant displacement of charge density from Pu towards the O and atomic H and the OH molecule is polarized. The dipoles formed on the O-Pu and H-Pu bonds will point towards the surface at an angle while the dipole formed on the upright O-H bond will point out of the surface. In case, a straightforward quantification of the alignment of induced net dipole moment is not possible and we rely on the work function to predict its orientation.

Table II shows the changes in the work function of the surface due to the presence of the adsorbates. The changes in the work function arise from the formation of surface dipoles due to the transfer of charge between the substrate and adsorbates. If the dipole moment points out of the surface then $\Delta\Phi < 0$. If the dipole moment points into the surface, then $\Delta\Phi > 0$. As argued earlier using the difference charge density plots, the work functions of the most favorable $T_1$ site for $H_2O$ should be negative as the dipole moment is oriented at about 10º with respect to surface ($H_2O$ transfer charge to the substrate). The negative work function is also true for the other case where the dipole orientation is nearly flat or out of the surface. If we take the opposite case of final $H_2O$ orientation -83º at the $B_2$ we clearly observe $\Delta\Phi > 0$ since the dipole points towards the surface (the substrate transfers charge to $H_2O$). Thus the work function enables us to gauge the various modes of charge transfer between the adsorbates and substrates.

In Figure 4, the angular momentum-resolved electronic density of states (EDOS) of the single particle Kohn-Sham energy eigenvalues for the free water molecule, clean surface, and water adsorbed on the surface at the most favorable $T_1$ site is shown. The three highest molecular orbitals of the $H_2O$, namely $1b_2$, $3a_1$, and $1b_1$, can clearly be seen. The $1b_2$ orbital is the σ orbital which mainly makes up the OH bonds, $3a_1$ orbital is also a σ orbital consisting mainly of oxygen lone pair electrons but with a small contribution to the OH bond, while $1b_1$ orbital is a π orbital consisting of oxygen lone pair electrons. Upon $H_2O$ adsorption, the $1b_2$, $3a_1$, and $1b_1$ peaks are stabilized to lower energies by 2.3 eV, 2.4 eV, and 2.7 eV respectively, implying that the $H_2O$ interacts with the surface mainly via the $1b_1$ orbital. This interaction explains why $H_2O$ adsorbs on the surface at the $T_1$ site in a nearly flat orientation because in such an orientation, the symmetry of the water molecule dictates that the $1b_1$ orbital lies nearly parallel to the surface normal and thus provide the maximum spatial overlap with the Pu electron states at the $T_1$ site. Another noticeable feature in Figure 4 is the character of the Pu-5$f$ density of states before and after adsorption. Clearly, a little or no change occurs in the shape of the Pu-5$f$ bands between the two plots, implying that the Pu-5$f$ electrons remain chemically inert in the adsorption process and that the interaction is governed by the delocalized Pu-6$d$ electron and the water $1b_1$ orbital.

In Figure 5, we depict the EDOS for the co-adsorbed dissociated products OH and H at the most stable $F_3+H_3$ sites. Specifically, the EDOS plots for the free OH molecule and H atom, the clean slab, OH and H co-adsorbed on the surface. In plot for free hydroxyl, the three highest molecular orbitals 3σ, 1π, and 4σ, as well the hydrogen 1$s$ orbital can clearly be seen. Upon adsorption, the OH 3σ and 1π orbitals and the H 1$s$ orbital are stabilized to lower energies, whereas the initially singly occupied OH 4σ orbital becomes an empty state. Also, a reduction in the Pu-5$f$ density of states below the Fermi level after adsorption in comparison to the clean slab can be observed, indicating that the some of Pu-5$f$ electrons participate in chemical bonding (Pu-to-adsorbate charge transfer).

Finally in Figure 6, we depict the EDOS for atomic O, atomic H, and 2H+O adsorbed at the most stable $F_3+F_3+F_3$ sites. First, the O $2p$ and H $1s$ states are stabilized to lower energies after adsorption in comparison to their respective atomic cases due to Pu-adsorbate charge transfer. Secondly, the O $2p$ and H $1s$ states hybridize with the Pu-$6d$ and Pu-$5f$ states. Thirdly, the Pu-$5f$ density of states at the Fermi level reduces after adsorption and the sharp peaks in the clean slab appear to broaden out, implying that some of Pu-$5f$ electrons become delocalized and thus participate in chemical bonding.

In summary, density functional theory has been used to study the interaction of water in molecular and dissociated states with the δ-Pu (111) surface. In the molecular state, water is physisorbed in an almost flat-lying orientation (nearly parallel to the surface) at site very close to the one-fold coordinated on-top site. The interaction of the water $1b_1$ orbital and the Pu-$6d$ orbital provides the stability of water on the surface, implying that the Pu-$5f$ electrons remain chemically inert. The co-adsorption of the partially dissociated (H+OH) and fully dissociated products at the three-fold hollow sites yielded chemisorption, coupled with the rumpling of the surface layer and delocalization of the Pu-$5f$ electrons. The dissociated products formed strong ionic bonds with the surface leading to a general decrease in the work function after chemisorption.

This work is supported by the Chemical Sciences, Geosciences and Biosciences Division, Office of Basic Energy Sciences, Office of Science, U. S. Department of Energy (Grant No. DEFG02-03ER15409) and the Welch Foundation, Houston, Texas (Grant No. Y-1525). This work also used resources of the National Energy Research Scientific Computing Center, supported by the Office of Science of the USDOE under

Contract No. DE-AC02-05CH11231, Texas Advanced Computing Center, Austin, Texas and the supercomputing facilities at the University of Texas at Arlington.

**References**

[1] L. R. Morss, N. M. Edelstein, J. Fuger (Eds.) and J. J. Katz (Hon. Ed.) *The Chemistry of the Actinide and Transactinide Elements, Vols. 1-5* (Springer, New York, 2006).

[2] D. K. Shuh, B. W. Chung, T. Albrecht-Schmitt, T. Gouder and J. D. Thompson (Eds), *Actinides 2008-Basic Science, Applications, and Technology*, Proceedings of the Materials Research Society, **1104** (2008).

[3] G. D. Jarvinen (Ed.) *Plutonium Futures˜ The Science*, American Institute of Physics Conference Proceedings, **673** (2003).

[4] A. M. Boring, J. L. Smith, Los Alamos Science, **1** (2000).

[5] D. Hoffman (Ed.) *Advances in Plutonium Chemistry 1967-2000* (American Nuclear Society, La Grange, Illinois and University Research Alliance, Amarillo, Texas, 2002).

[6] M. J. Fluss, D. E. Hobart, P. G. Allen, J. D. Jarvinen (Eds.) *Proceedings of the Plutonium Futures – The Science 2006 Conference*, J. Alloys. and Comp. **444-445** (2007).

[7]  M. A. Henderson, Surf. Sci. Rep. **46** (2002) 1.

[8] J. M. Haschke, T. H. Allen, J. L. Stakebake, J. Alloys Comp. **243** (1996) 23.

[9] J. C. Boettger, A. K. Ray, Int. J. Quant. Chem. **90** (2002) 1470.

[10] X. Wu, A. K. Ray, Phys. Rev. B **65** (2002) 085403.

[11] P. J. Hay, Mat. Res. Symp. Proc. **893** (2006) 0893-JJ08-04.1.

[12] P. Perdew, K. Burke, M. Ernzerhof, Phys. Rev. Lett. **77**, (1996) 3865.

[13] B. Delley, J. Chem. Phys. **92** (1990) 508.

[14] B. Delley, J. Chem. Phys. **113** (2000) 7756.

[15] P. Blaha, K. Schwarz, G. K. H. Madsen, D. Kvasnicka, J. Luitz, *WIEN2k, An Augmented Plane Wave Plus Local Orbitals Program for Calculating Crystal properties* (Vienna University of Technology, Austria, 2001).

[16] D. D. Koelling and B. N. Harmon, J. Phys. C, **10** (1977) 3107.

[17] P. A. Thiel, T. E. Madey, Surf. Sci. Rep. **7** (1978) 211.

[18] R. Atta-Fynn, A. K. Ray, Phys. Rev. B **75** (2007) 195112.

[19] A. Michaelides, V. A. Ranea, P. L. de Andres, D. A. King, Phys. Rev. B **69** (2004) 075409.

[20] M. N. Huda, A. K. Ray, Phys. Rev. B **72** (2005) 085101.

Table I: Optimized geometric parameters and adsorption energies of $H_2O$, OH+H, and O+2H, adsorbed on the δ-Pu (111) surface. $E_{ads}$ is the adsorption energy per adsorbate, the angle α defined in Figure 1 (c) is used to label the initial and final states $α_i$ and $α_f$, $\Delta O_{xy}$ is the lateral deviation of the O part $H_2O$ from the precise final adsorption site, $\Delta_{12}$ is the interlayer relaxation of the surface layer with respect to the subsurface layer, $\Delta r$ is the rumpling of the surface layer (maximum vertical displacement between atoms on that layer), θ is the HOH bond angle, $d_{XY}$ is the average XY bond length, and N is the adatom coordination number based on a nearest neighbor cut-off of 3 Å. The adsorption sites A+B corresponding to the dissociated products X+Y are such that X occupies site A and Y occupies site B.

| $H_2O$ adsorption | | | | | | | | | | | |
|---|---|---|---|---|---|---|---|---|---|---|---|
| $E_{ads}$ (eV) | Initial state $α_i$ | Initial site | Final state $α_f$ | Final Site | $\Delta O_{xy}$ (Å) | $\Delta_{12}$ (%) | $\Delta r$ (Å) | $d_{OH}$ (Å) | θ | $d_{OPu}$ (Å) | $d_{HPu}$ (Å) | N |
| 0.59 | 0° | $T_1$ | 10° | $T_1$ | 0.21 | 2.0 | 0.33 | 0.96 | 105° | 2.63 | 2.93 | 3 |
| 0.59 | 0° | $B_2$ | 12° | $T_1$ | 0.09 | 1.7 | 0.20 | 0.96 | 105° | 2.63 | 2.93 | 3 |
| 0.53 | 0° | $H_3$ | -2° | $T_1$ | 0.79 | 2.4 | 0.64 | 0.96 | 104° | 2.64 | 2.94 | 3 |
| 0.54 | 0° | $F_3$ | -3° | $T_1$ | 0.79 | 2.3 | 0.65 | 0.97 | 104° | 2.65 | 2.94 | 3 |
| 0.58 | 90° | $T_1$ | 9° | $T_1$ | 0.30 | 2.1 | 0.38 | 0.96 | 105° | 2.63 | 2.95 | 3 |
| 0.35 | 90° | $B_2$ | 6° | $B_2$ | 0.39 | 1.8 | 0.08 | 0.96 | 103° | 3.23 | 3.50 | |
| 0.42 | 90° | $H_3$ | 52° | $B_2$ | 0.48 | 2.4 | 0.67 | 0.96 | 108° | 2.86 | 3.59 | 2 |
| 0.43 | 90° | $F_3$ | 61° | $B_2$ | 0.31 | 2.5 | 0.61 | 0.97 | 108° | 2.85 | 3.34 | 2 |
| 0.59 | -90° | $T_1$ | 9° | $T_1$ | 0.26 | 2.2 | 0.34 | 0.96 | 105° | 2.63 | 2.93 | 3 |
| 0.57 | -90° | $B_2$ | 6° | $T_1$ | 0.43 | 2.3 | 0.50 | 0.96 | 105° | 2.64 | 2.95 | 3 |
| 0.26 | -90° | $H_3$ | -84° | $B_2$ | 0.01 | 1.5 | 0.05 | 0.96 | 104° | 3.63 | 3.20 | |
| 0.17 | -90° | $F_3$ | -83° | $B_2$ | 0.03 | 1.4 | 0.05 | 0.96 | 104° | 3.62 | 3.25 | |

| (OH+H) co–adsorption | | | | | | | | | | |
|---|---|---|---|---|---|---|---|---|---|---|
| $E_{ads}$ (eV) | Initial state $α_i$ | Initial site | Final state $α_f$ | Final Site | $\Delta_{12}$ (%) | $\Delta r$ (Å) | $d_{OH}$ (Å) | $d_{OPu}$ (Å) | $d_{HPu}$ (Å) | N |
| 5.19 | 90° | $H_3+H_3$ | 90° | $H_3+H_3$ | 7.1 | 0.32 | 0.96 | 2.44 | 2.25 | 6 |
| 5.16 | 90° | $F_3+F_3$ | 90° | $F_3+F_3$ | 2.8 | 0.93 | 0.96 | 2.43 | 2.25 | 6 |
| 5.19 | -90° | $H_3+H_3$ | 90° | $H_3+H_3$ | 7.0 | 0.20 | 0.95 | 2.44 | 2.26 | 6 |
| 5.18 | -90° | $F_3+F_3$ | 90° | $H_3+F_3$ | 5.9 | 0.71 | 0.96 | 2.43 | 2.25 | 6 |

| (O+H+H) co-adsorption | | | | | |
|---|---|---|---|---|---|
| $E_{ads}$ (eV) | Initial Sites | $\Delta_{12}$ (%) | $\Delta r$ (Å) | $d_{OPu}$ (Å) | $d_{HPu}$ (Å) | N |
| 5.46 | $H_3+H_3+H_3$ | 0.5 | 1.0 | 2.22 | 2.22 | 9 |
| 5.51 | $F_3+F_3+F_3$ | 0.1 | 1.5 | 2.16 | 2.20 | 9 |

Table II: Change in the work function ΔΦ for each slab-with-adsorbate system. ΔΦ = Φ(Pu slab+adsorbate) - Φ(Pu slab), where Φ(Pu slab) is the work function of the clean surface and Φ(Pu slab+adsorbate) is the work function of the surface in the presence of the adatom. Φ(Pu slab) = 3.28 eV.

|  | Final state $\alpha_f$ | Final Site | ΔΦ (eV) |
|---|---|---|---|
| $H_2O$ | 10° | $T_1$ | -0.35 |
|  | 12° | $T_1$ | -0.35 |
|  | -2° | $T_1$ | -0.15 |
|  | -3° | $T_1$ | -0.13 |
|  | 9° | $T_1$ | -0.34 |
|  | 6° | $B_2$ | -0.08 |
|  | 52° | $B_2$ | -1.20 |
|  | 61° | $B_2$ | -1.20 |
|  | 9° | $T_1$ | -0.32 |
|  | 6° | $T_1$ | -0.31 |
|  | -84° | $B_2$ | 1.34 |
|  | -83° | $B_2$ | 1.26 |

|  | Final state $\alpha_f$ | Final Site | ΔΦ (eV) |
|---|---|---|---|
| OH+H | 90° | $H_3+H_3$ | -0.60 |
|  | 90° | $F_3+F_3$ | -0.80 |
|  | 90° | $H_3+H_3$ | -0.60 |
|  | 90° | $H_3+F_3$ | -0.60 |

|  | Initial Sites |  | ΔΦ (eV) |
|---|---|---|---|
| O+H+H | $H_3+H_3+H_3$ |  | -0.06 |
|  | $F_3+F_3+F_3$ |  | 0.24 |

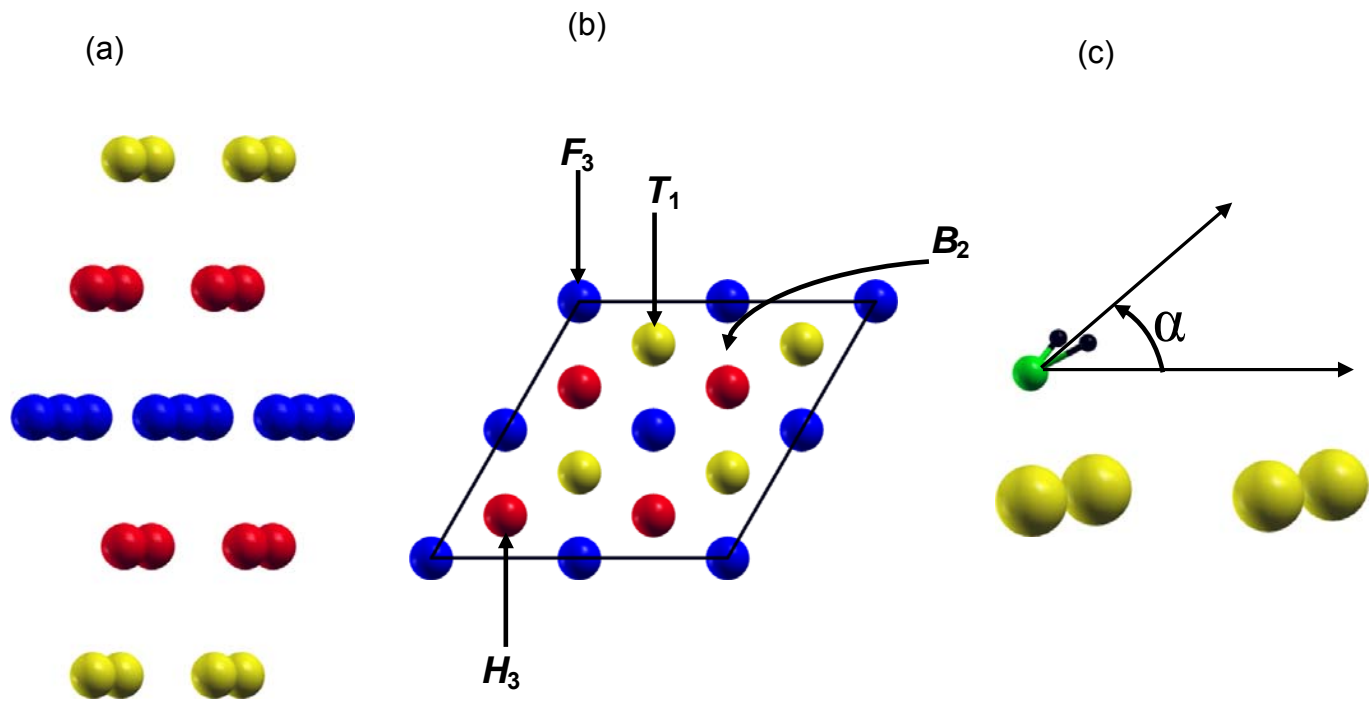

Figure 1:
(a) Side view of clean five-layer symmetric δ-Pu (111) slab with the surface layer, subsurface layer, and bulk layer Pu atoms colored gold, red, and blue, respectively.
(b) Top view of the clean δ-Pu (111) slab with adsorption site labels: one-fold on-top site $T_1$ (adsorbate is directly on top of a surface Pu atom on the surface layer); two-fold bridge site $B_2$ (adsorbate is placed in the middle of a Pu-Pu bond); three-fold hollow hcp site $H_3$ (admolecule sees a Pu atom located on the subsurface layer); and iv) three-fold hollow fcc site $F_3$ (admolecule sees a Pu atom on the bulk layer).
(c) An illustration of the angle α defined as the angle between the molecular dipole plane and the unrumpled surface. Thus α = 90° corresponds to a vertical upright molecule with the H atoms pointing up (away from the surface), α = -90° corresponds to a vertical upright molecule with the H atoms pointing down (towards the surface), and α = 0° corresponds to a flat-lying molecule (parallel to the surface).

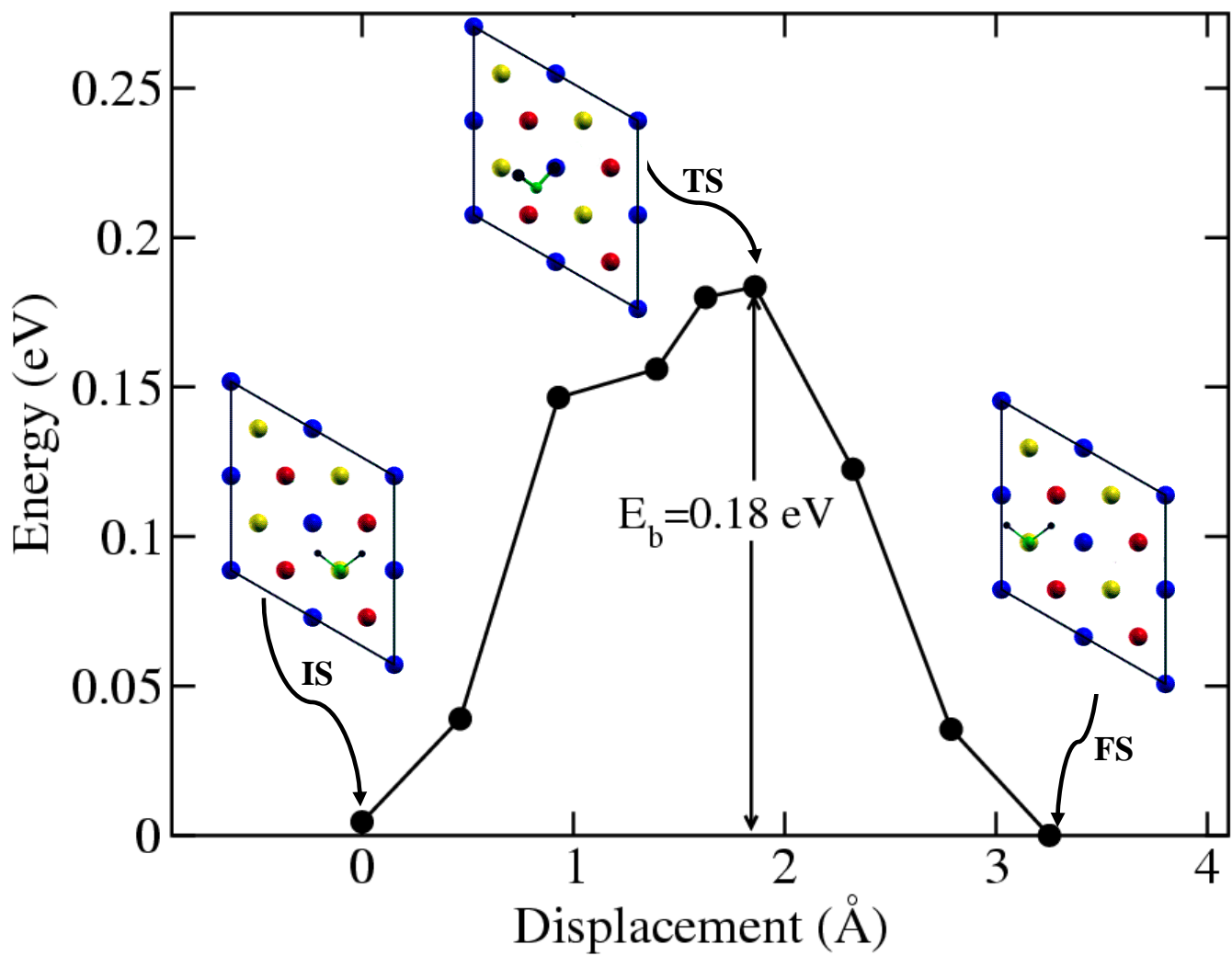

Figure 2: Variations in the $H_2O/\delta$-Pu(111) total energy as a function of the lateral displacement of O part of the molecule from an on-top $T_1$ site to a neighboring on-top $T_1$ site via the $T_1$-$B_2$-$T_1$ diffusion pathway. IS, TS, and FS denotes the initial, transition, and final states respectively. The lowest energy state is set to zero and the activation energy for the diffusion is denoted by $E_b$. The solid lines connecting the data points serve as a guide to the eye. Atom coloring scheme is the same as in Figure 1(a).

(a) H$_2$O/δ-Pu(111) at the $T_1$ site

(b) (OH+H)/δ-Pu(111) at the $F_3$+$H_3$ sites

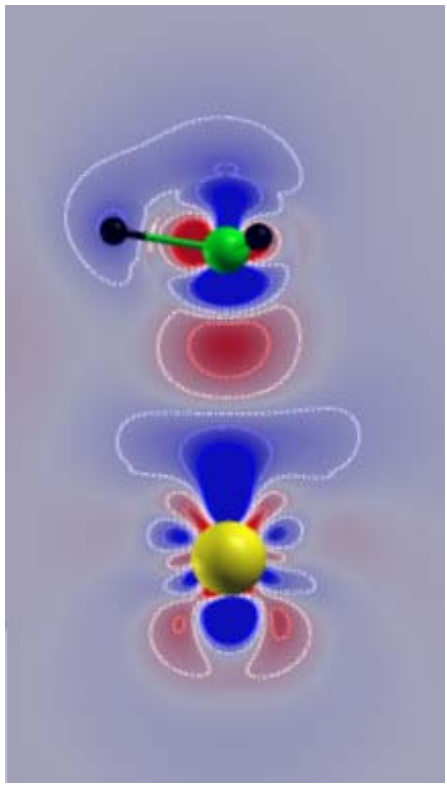
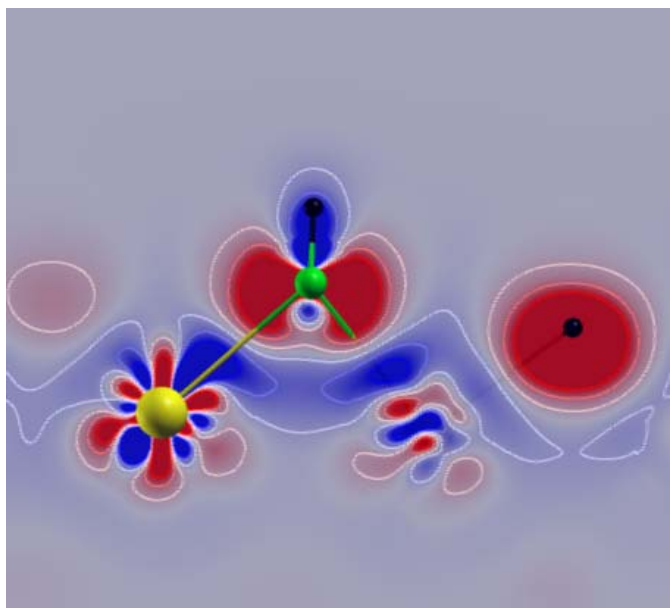

(c) (O+H+H)/δ-Pu(111) at the $F_3$+$F_3$+$F_3$ sites

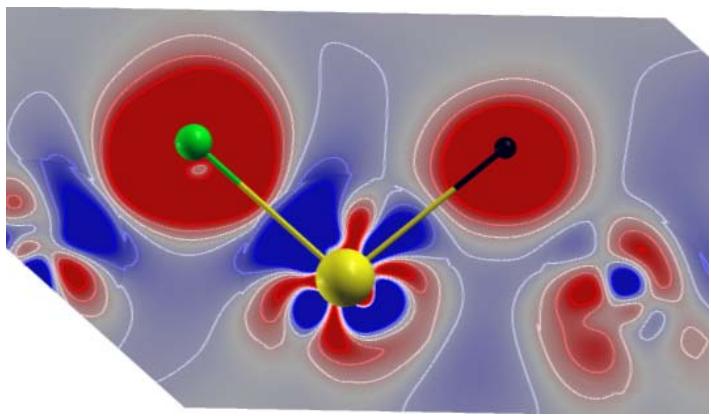
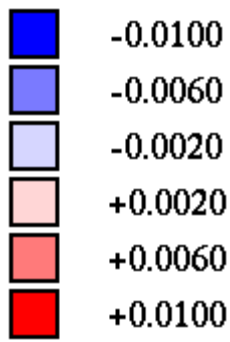

Figure 3: Difference charge density plots for the most stable cases corresponding to (a) molecular adsorption, (b) partially dissociated adsorption, and (c) fully dissociated adsorption configuration. Regions of excess charge and charge depletion are colored red and blue respectively. Atom coloring scheme is the same as in Figure 1(a).

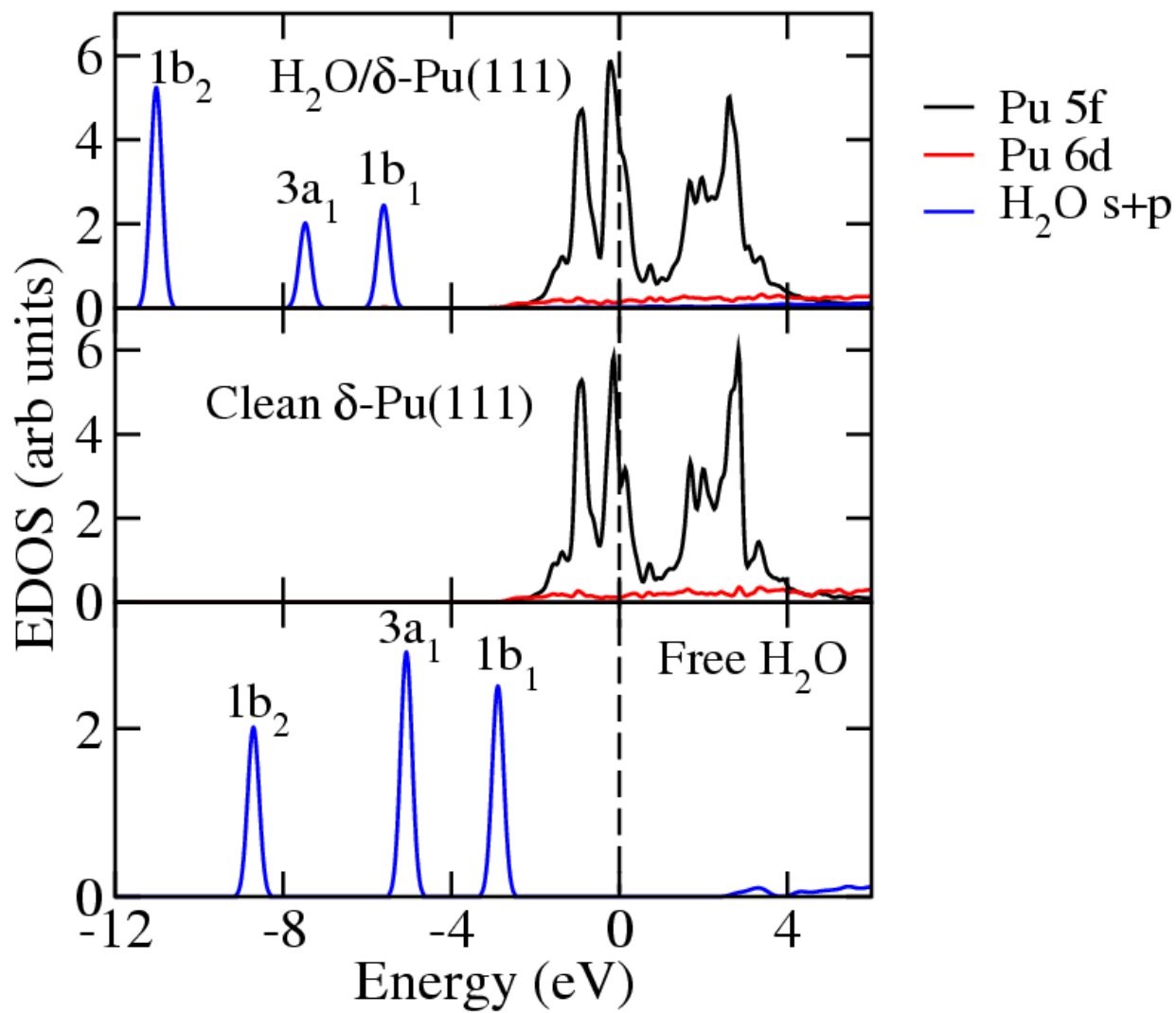

Figure 4: Partial density of states the free water molecule, clean slab, and water adsorbed on the δ-Pu(111) surface at the most stable $T_1$ site. The dashed vertical line is the Fermi level.

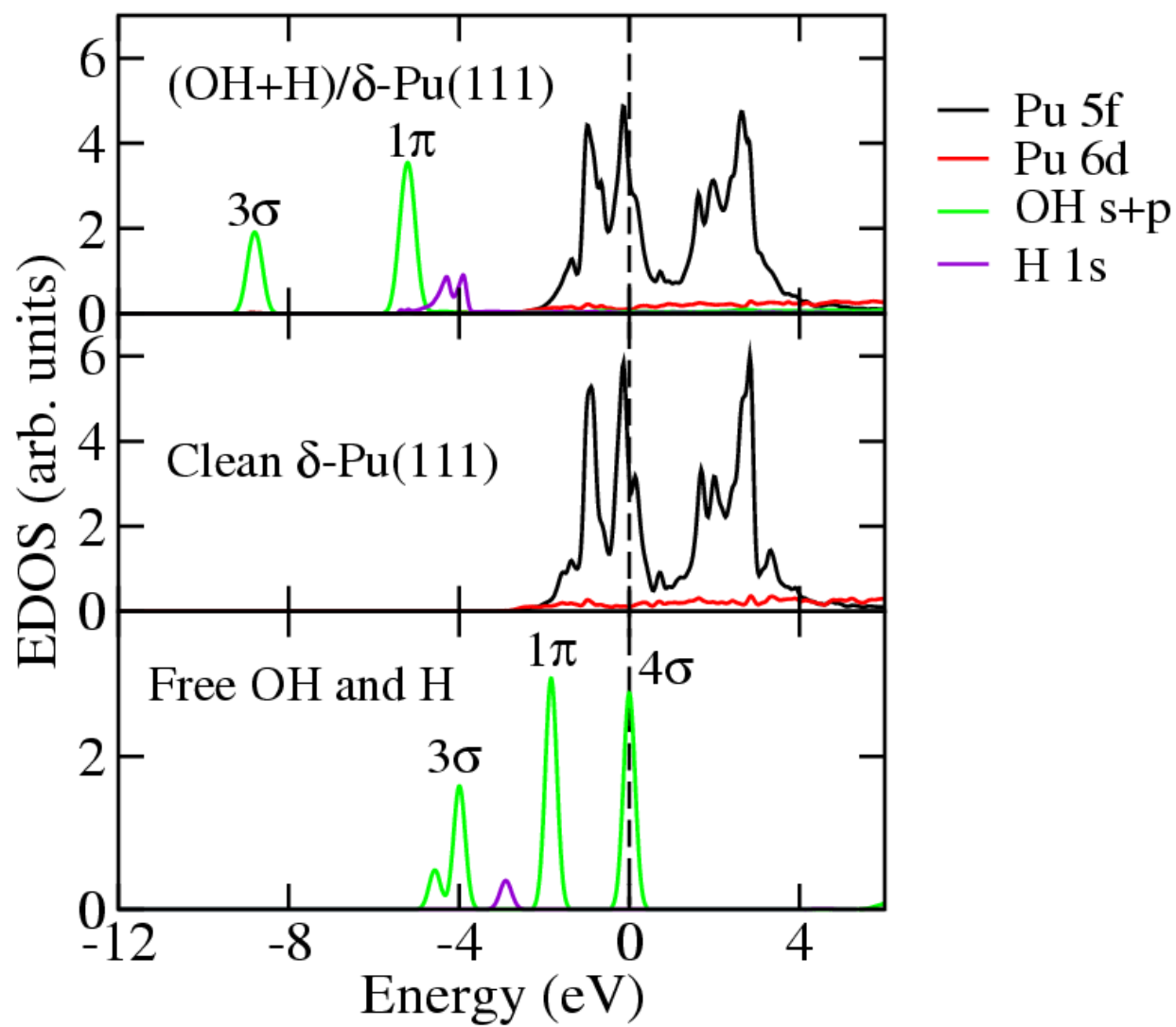

Figure 5: Partial density of states for free H, free OH clean slab, and OH and H co-adsorbed on δ-Pu(111) at the $F_3+H_3$ sites. The dashed vertical line is the Fermi level.

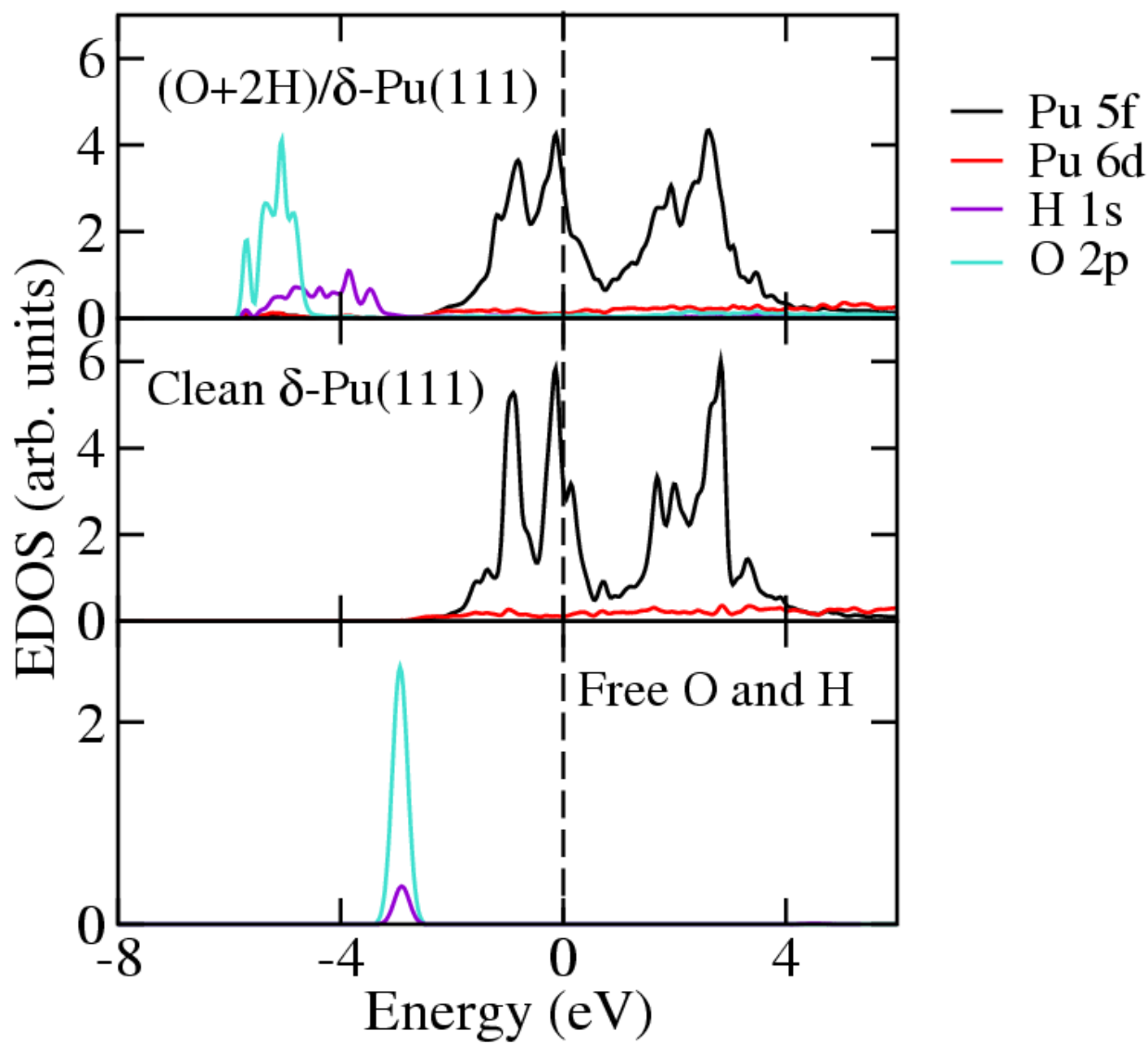

Figure 6: Partial density of states for free H, free O, clean slab, and O and two H atoms co-adsorbed on δ-Pu(111) at the most stable $F_3+F_3+F_3$ sites. The dashed vertical line is the Fermi level.